\documentclass{PoS}
\usepackage[outercaption]{sidecap}
\usepackage{slashed}
\usepackage{latexsym,bm}
\usepackage{amsmath}
\usepackage{amssymb}
\usepackage{color}

\newcommand{\Order}{\mathcal{O}}

\title{Another walk through the world of chiral dynamics}

\ShortTitle{Another walk through the world of chiral dynamics}

\author{\speaker{Ulf-G. Mei{\ss}ner}\thanks{Work supported in part by
         DFG \& NSFC (CRC 110), CAS PIFI (2018DM0034), and VolkswagenStiftung (93652) }\\
        Helmholtz-Institut f\"ur Strahlen- und Kernphysik and Bethe
        Center for Theoretical Physics
        Universit\"at Bonn,  D-53115 Bonn, Germany \\
        Institute for Advanced Simulation, Institut f\"{u}r Kernphysik
        and J\"ulich Center for Hadron Physics,
        Forschungszentrum J\"{u}lich,
        D-52425 J\"{u}lich, Germany \\
        Tbilisi State  University,  0186 Tbilisi, Georgia\\
        E-mail: \email{meissner@hiskp.uni-bonn.de}}

\abstract{Chiral dynamics is a pretty mature field. Nonetheless, there are many exciting new
  developments. In this opening talk, I consider S-wave, isospin-zero pion-pion scattering and
  the calculation of the width of the lightest baryon resonances at two loops. New insights into
  the chiral dynamics of  charm-strange mesons are discussed as well as recent results on the
  flavor decomposition of the pion-nucleon $\sigma$-term. I end with a short wish-list of lattice
  QCD tests pertinent to chiral dynamics.
               }

\FullConference{The 9th International Workshop on Chiral Dynamics,\\
		September 17 -21, 2018\\
		Duke University, Durham, North Carolina, USA}

\begin{document}

\section{Introduction}

Chiral dynamics is a pretty mature field. We have high-precision calculations and tests in the
$\pi\pi$ and $\pi N$ systems, that is in the two-flavor sector of the light up and down quarks.
This is largely based on a fruitful marriage of chiral perturbation theory
(CHPT) with dispersion theory. In case of pion-pion scattering, such a scheme also allows one
to pin down precisely the parameters of the lightest resonances in QCD. There is also much
progress in lattice QCD. However, especially in the
sector with baryon  number $B\geq 1$, there is still a lack of calculations of scattering processes,
not  even to talk about precision. The situation in the three-flavor sector remains ambiguous as
$m_s \sim \Lambda_{\rm QCD}$. Still, there is some marked progress  in eta and kaon decays.
When it comes to baryons, the most urgently needed calculation concerns the flavor-singlet piece
of the pion-nucleon $\sigma$-term, $\sigma_0$, as I will discuss later. In this context, the lattice can and
does provide numbers on the strange matrix element $m_s \langle p|\bar{s}s|p\rangle$, which nicely
complements the dispersive determination of $\sigma_{\pi N}$. Furthermore, unitarization schemes
offer insight into excited states, though at some cost. In often used approximations, the analytical
structure is not always respected, so improvements of the method for coupled channels are required.
Nevertheless, there has been some remarkable progress in heavy-light systems combining unitarized
CHPT, lattice QCD results and accurate data.
In what follows, I give more details on some of these statements but the reader is advised to 
consult the many fine talks given at this workshop for further discussions and references.

\section{Lesson 1: Once more on the $\pi\pi$ scattering length $a_0$}

\begin{figure}[t]
\begin{center}
$\underbrace{\includegraphics[height=1.1cm]{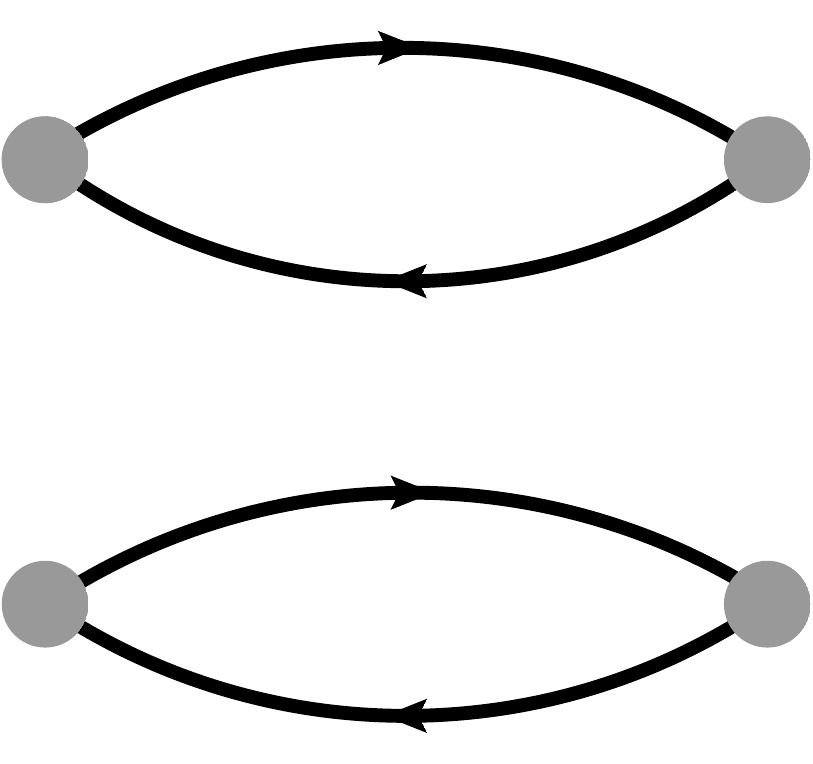}~~
\includegraphics[height=1.1cm]{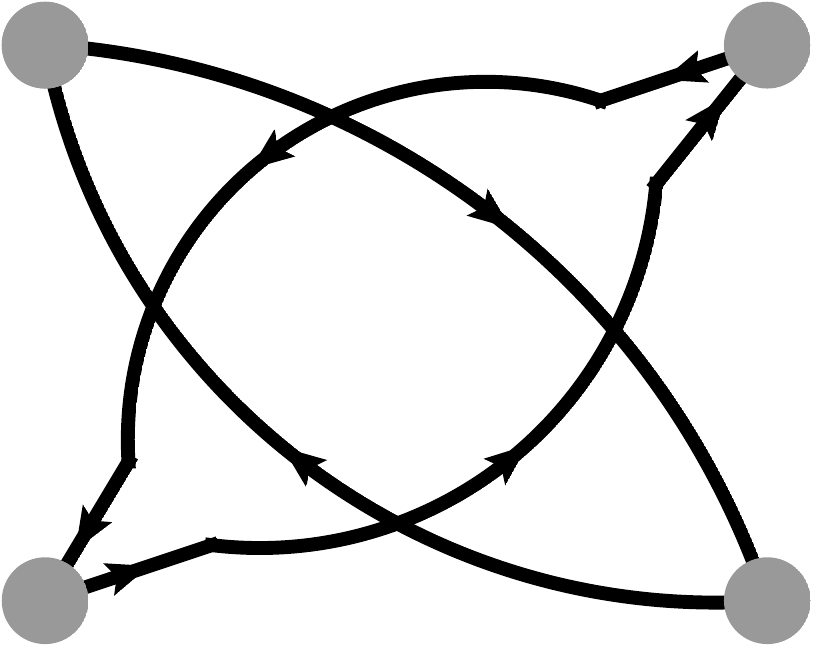}}_{\rm Direct}
~~~~~~\underbrace{\includegraphics[height=1.1cm]{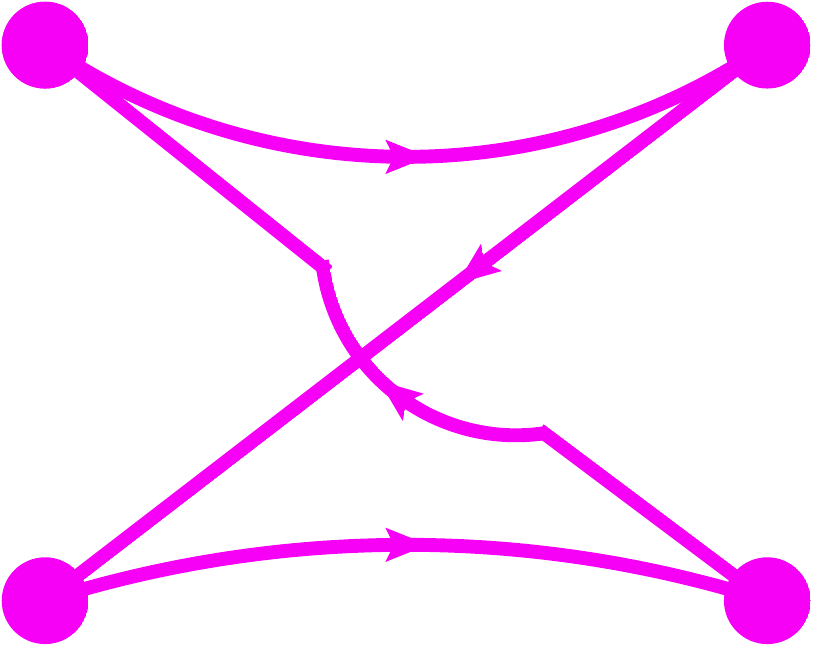}}_{\rm Crossed}
~~~~~~\underbrace{\includegraphics[height=1.1cm]{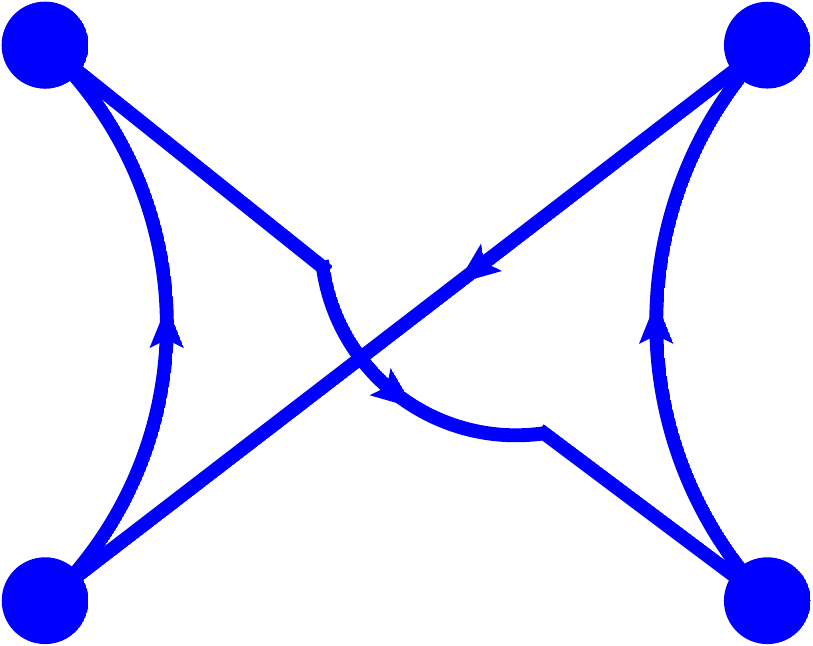}~~
\includegraphics[height=1.3cm]{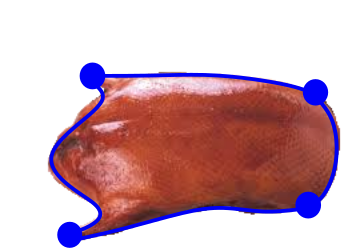}}_{\rm Rectangular}
~~~~~~\underbrace{\includegraphics[height=1.2cm]{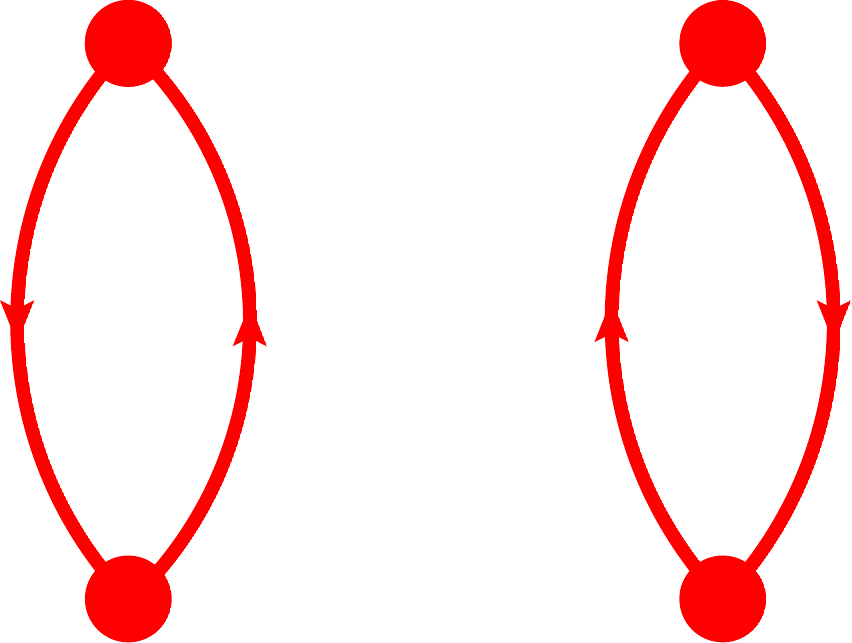}}_{\rm Vacuum}$
\caption{Types of Wick contractions in pion-pion scattering.}
\label{fig:wick}
\end{center}
\vspace{-6mm}
\end{figure}

 Elastic pion-pion scattering ($\pi\pi\to\pi\pi$) is the purest process 
in two-flavor chiral dynamics as the up and the down quark masses are
really small compared to any other hadronic mass scale. At threshold 
the scattering amplitude is given in terms of two numbers, the 
scattering lengths $a_0$ and $a_2$. In my 2012 talk \cite{Meissner:2012ku},
I  had reviewed the experimental and the theoretical status, concluding:
``This is truly  one of the finest tests of the Standard Model at low energies.
However, not all is well --  a direct lattice determination of $a_0$ is still
missing and the lattice practitioners are urged to provide this so important
number. Such a calculation is, of course, technically challenging because
of the disconnected diagrams, but time is ripe for doing it.'' Since then,
only two groups \cite{Fu:2013ffa,Liu:2016cba} have published results on this so important quantity, summarized
in Tab.~\ref{tab:pipi}. Both of these are consistent with the high-precision
result from Roy equations, $a_0 = 0.220\pm 0.005$~\cite{Colangelo:2000jc}, but presumably
underestimate the systematic errors. There have also been very nice lattice calculations
concerning the S-wave, isospin-zero $\pi\pi$ phase shifts and the $f_0(500)$ (sigma)
meson~\cite{Briceno:2016mjc,Guo:2018zss}, but these do not give values for $a_0$.
  \begin{table}[h]
\begin{center}  
\begin{tabular}{|lccc|}
\hline
Author(s) & $a_0$         & Fermion action & Pion mass range \\ 
\hline
Fu \cite{Fu:2013ffa}         & 0.214(4)(7)   &  asqtad staggered& 240 - 430 MeV\\
Liu et al.\cite{Liu:2016cba}  & 0.198(9)(6)  &  twisted mass  & 250 - 320 MeV\\
\hline
\end{tabular}
\caption{Lattice QCD determinations of $a_0$.}
\label{tab:pipi}.
\end{center}
\vspace{-9mm}
\end{table}
\noindent
To get a better handle on the disconnected diagrams, connected and disconnected contractions in $\pi\pi$
scattering have been analyzed in Ref.~\cite{Acharya:2017zje}. There are four types of contractions, namely
the connected ones, called direct (D) and crossed (C) contractions, the singly disconnected ones of rectangular
(or roasted Peking duck\footnote{As introduced by Feng-Kun Guo in 2017, see also Fig.~\ref{fig:wick}
that was generated by him.}) type (R) and the doubly disconnected vacuum graphs (V), which are most difficult to
calculate in lattice QCD, cf. Fig.~\ref{fig:wick}. By means of partially-quenched chiral perturbation (PQCHPT)
theory~\cite{Bernard:1993sv,Sharpe:2000bc,Sharpe:2001fh,Bernard:2013kwa},
one can  distinguish and analyze the effects from different types of contraction diagrams to the pion-pion
scattering amplitude, in particular also to $a_0$. The findings of Ref.~\cite{Acharya:2017zje} can be summarized
as follows: The R and C-type diagrams are leading order (LO) in the chiral and the large-$N_C$ expansion, see
also~\cite{Guo:2013nja}, whereas the D- and V-type diagrams are next-to-leading order (NLO) in the chiral and the
large-$N_C$ expansion. As expected, the R-type diagram dominates when it contributes
(as is the case for $a_0$), while neglecting the V-type contribution reduces $a_0$ by about 12\%.

This program was recently carried further in Ref.~\cite{Acharya:2019meo}. One has to make a
direct connection to the energy levels measured in lattice QCD. This can be achieved in the following way:
First, each contraction in $\pi\pi$ scattering can be represented by a physical scattering process
between two pions in SU$(4|2)$ partially quenched QCD.
Then, the multi-channel scattering matrix can be diagonalized to obtain four effective single-channel
scattering amplitudes. Two of these (labeled $\alpha, \beta$) involve only connected contractions
and the others (called $\gamma, \delta$) contain
disconnected pieces. With that, one can immediately relate the threshold parameters of each single-channel
scattering amplitude to the corresponding discrete energy levels in a finite volume through the
usual single-channel L\"uscher formula. Further, the threshold parameters can be expressed in terms of physical
and unphysical LECs in the SU$(4|2)$ PQCHPT, which can be fitted to discrete energy levels extracted
from the connected $\pi\pi$ correlation functions. In particular, fitting to certain data form the
European Twisted Mass Collaboration~\cite{Helmes:2015gla},  one achieves an order-of-magnitude
improvement over the earlier result~\cite{Boyle:2015exm} in the determination of the LECs combination
$3L_0^{\mathrm{PQ},r}+L_3^{\mathrm{PQ},r}.$ Performing lattice computations using more volumes,
one can also improve the precision of the worst known LEC $L_0^{\text{PQ},r}.$ With these fitted
LECs one is able to predict the discrete energy shifts $\delta E_0^{\gamma,\delta}$, which involve
disconnected contractions, as functions of the lattice size. For more details, see~\cite{Acharya:2019meo}.

\section{Lesson 2: The width of the lightest baryon resonances}

In this section, I will be concerned with calculating the width of the two lightest
baryon resonances, the $\Delta(1232)$ and the Roper $N^*(1440)$. This might
at first sight appear irritating, as imaginary parts are usually not precisely
reproduced in CHPT. For that simple reason, one has to employ a complex-mass
scheme and work to two loops.

\begin{figure}[t]
\centering
\includegraphics[width=0.84\textwidth]{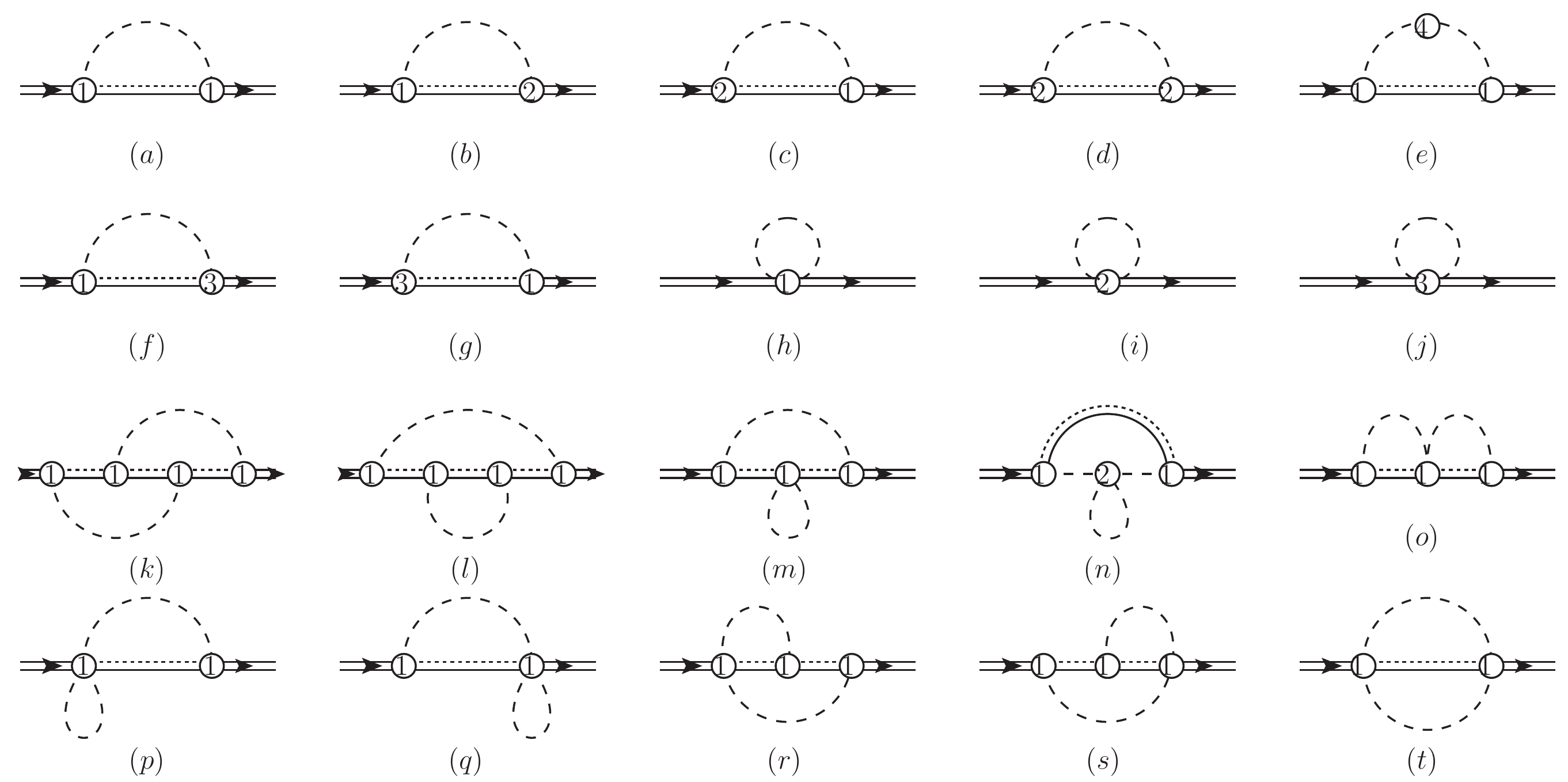}~~~~~
\caption{One and two-loop self-energy diagrams contributing to the width 
of the delta resonance up-to-and-including 
fifth order according to the standard power counting. The dashed and double solid lines 
represent the pions and the delta resonances, respectively. 
The double (solid-dotted) lines in the loops correspond to either nucleons or deltas. 
The numbers in the circles give the chiral orders of the vertices.}
\label{fig:Delta}
\vspace{-0mm}
\end{figure}

\begin{figure}[t]
\centering
\includegraphics[width=0.59\textwidth]{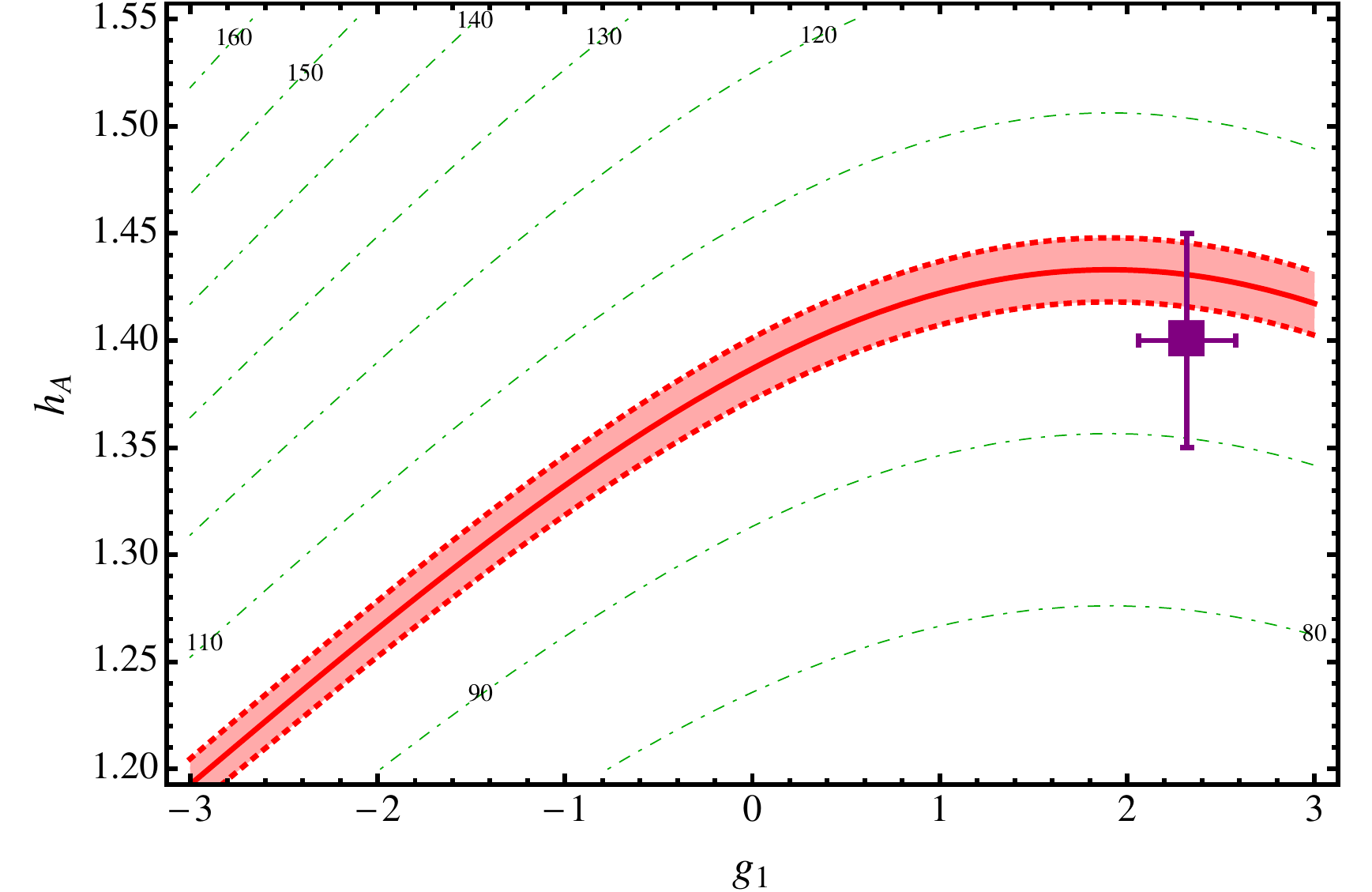}
\caption{Correlation between the leading $\pi \Delta$ and $\pi N \Delta$ couplings. The band gives the 
uncertainty of the calculation, and the box with the error bars are the results from the analysis of
Ref.~\cite{Siemens:2016jwj}.}
\label{fig:corr}
\vspace{-3mm}
\end{figure}

Consider first the width of the $\Delta$ at two-loop order~\cite{Gegelia:2016pjm}.
The  pertinent effective Lagrangian contains, besides many
other terms, the leading $\pi \Delta$ and $\pi N\Delta$ couplings,
parameterized in terms of the LECs $g_1$ and $h$, respectively,
\begin{eqnarray}
{\cal L}^{(1)}_{\pi\Delta} &=& -\bar{\Psi}_{\mu}^i\xi^{\frac{3}{2}}_{ij}\Bigl\{\left(i\slashed{D}^{jk}
-m_{\Delta}\delta^{jk}\right)g^{\mu\nu}
-i\left(\gamma^\mu D^{\nu,jk}+\gamma^\nu D^{\mu,jk}\right) +i \gamma^\mu\slashed{D}^{jk}\gamma^\nu\nonumber\\
%\end{eqnarray}
%\begin{eqnarray}
&+&m_{\Delta}\delta^{jk} \gamma^{\mu}\gamma^\nu
+{\color{red}g_1}\frac{1}{2}\slashed{u}^{jk}\gamma_5g^{\mu\nu}+g_2\frac{1}{2} (\gamma^\mu u^{\nu,jk}
+u^{\nu,jk}\gamma^\mu)\gamma_5 + g_3\frac{1}{2}\gamma^\mu\slashed{u}^{jk}\gamma_5\gamma^\nu \Bigr\}\xi^{\frac{3}{2}}_{kl}
{\Psi}_\nu^l~,\nonumber\\
{\cal L}^{(1)}_{\pi N\Delta} &=& {\color{red}h}\,\bar{\Psi}_{\mu}^i\xi_{ij}^{\frac{3}{2}}
\Theta^{\mu\alpha}(z_1)\ \omega_{\alpha}^j\Psi_N+ {\rm h.c.}~,
\label{eq:delta}
\end{eqnarray}  
in terms of the spin-1/2 (3/2) fields $\Psi_N~(\Psi_\mu^i)$ and for other notations, see~\cite{Gegelia:2016pjm}.
The power counting  rests on $m_\Delta - m_N$ being a small quantity. However, there are so many LECs (not all
are shown in Eq.~(\ref{eq:delta})), so how can one one possibly make a prediction? The trick is to use the
complex-mass renormalization scheme, a method that was originally introduced for precision $W,Z$-physics,
see e.g.~\cite{RGS,Denner:1999gp}  and later transported to chiral EFT~\cite{Djukanovic:2009zn}. For the
case under consideration, we need to evaluate the $\Delta$ self-energy on the complex pole,
\begin{equation}
   z - m_{\Delta}^0 - \Sigma(z) =0 ~~~~~~~~~{\rm with}~~~~~~~~~~z=m_\Delta-i\,\displaystyle\frac{\Gamma_\Delta}{2}~.
\end{equation}
The corresponding diagrams for the one- and two-loop self-energy contributing to the 
width of the delta resonance up to order $q^5$ are  displayed in Fig.~\ref{fig:Delta}, where the
counterterm diagrams are not shown. The one-loop diagrams are easily obtained, for the calculation
of the two-loop graphs one uses the Cutkovsky rules for instable particles, that relate the width to
the pion-nucleon scattering amplitude, $\Gamma_\Delta \sim |A(\Delta\to N\pi)|^2$~\cite{Veltman:1963th}.
One finds a remarkable  reduction of parameters that is reflected in the relation
\begin{eqnarray}
h_A &=& h - \left(b_3\Delta_{23}+b_8\,\Delta_{123}\right)
-\left(f_1\Delta_{23}+f_2\,\Delta_{123}\right)\Delta_{123}+2(2f_4-f_5)M_\pi^2~,\nonumber\\
\Delta_{23} &=&m_N-m_\Delta,  ~~~~~~\Delta_{123}=(M_\pi^2+m_N^2-m_\Delta^2)/(2m_N)~,
\end{eqnarray}  
which means that all of the LECs appearing in the $\pi N\Delta$ interaction at second
and third order, the $b_i (i=3,8)$ and $f_i (i=1,2,4,5)$, respectively, merely lead to a renormalization
of the LO $\pi N\Delta$ coupling $h$,
%\begin{equation}
%h_A = h - \left(b_3\Delta_{23}+b_8\,\Delta_{123}\right)
%-\left(f_1\Delta_{23}+f_2\,\Delta_{123}\right)\Delta_{123}+2(2f_4-f_5)M_\pi^2~,
%\end{equation}
and, consequently, one finds a very simple formula for the decay width $\Delta \to N\pi$,
\begin{equation}
\Gamma(\Delta\to N \pi) =  (53.91\,{h}_A^2+0.87g_1^2 {h}_A^2-3.31g_1^{} {h}_A^2 
-0.99\,{h}_A^4)~{\rm MeV}~.
\end{equation}
This leads to a novel correlation, that is independent of the number of colors, as depicted in
Fig.~\ref{fig:corr}. It is obviously fulfilled by the analyis of  Ref.~\cite{Siemens:2016jwj}, that showed that 
the inclusion of the $\Delta$ alleviates the tension between the threshold and subthreshold regions in the
description of $\pi N$ scattering found in baryon CHPT.

Next, I consider the calculation of the width of the Roper $N^*(1440)$ at two-loop order~\cite{Gegelia:2016xcw}.
A remarkable feature of the Roper is the fact that its decay width into a nucleon and a pion is similar to the width
%$\Gamma (R\to N\pi) \simeq \Gamma (R\to N\pi\pi) $
into a nucleon and two pions. Any model that is supposed to describe the Roper must account for this fact. In CHPT,
consider the  effective chiral Lagrangian of pions, nucleons and
deltas~\cite{Borasoy:2006fk,Djukanovic:2009gt,Long:2011rt},
\begin{eqnarray}
{\cal L}_{\rm eff}&=&{\cal L}_{\pi\pi}+{\cal L}_{\pi N}+{\cal L}_{\pi \Delta}+{\cal L}_{\pi R}
+{\cal L}_{\pi N\Delta}+{\cal L}_{\pi NR}+{\cal L}_{\pi\Delta R}~,\nonumber\\
{\cal L}_{\pi R}^{(1)}&=&\bar{\Psi}_R\left\{i\slashed{D}-m_R+\frac{1}{2}{\color{red}g_R}\slashed{u}
\gamma^5\right\}\Psi_R~,\quad
{\cal L}_{\pi R}^{(2)} = \bar{\Psi}_R\left\{c_1^R\langle\chi_+\rangle\right\}\Psi_R + \ldots~,\nonumber\\
{\cal L}_{\pi NR}^{(1)} &=& \bar{\Psi}_R\left\{\frac{1}{2}{\color{red}g_{\pi NR}}\gamma^\mu\gamma_5 u_\mu\right\}\Psi_N
+ {\rm h.c.}~, \quad\!\!\!\!
{\cal L}^{(1)}_{\pi \Delta R} = {\color{red}h_R}\,\bar{\Psi}_{\mu}^i\xi_{ij}^{\frac{3}{2}}\Theta^{\mu\alpha}(\tilde{z})\ \omega_{\alpha}^j\Psi_R+ {\rm h.c.}~,
\end{eqnarray}
where the leading Roper-pion, Roper-nucleon pion and Delta-Roper-pion couplings, $g_R$, $g_{\pi NR}$ and
$h_R$, respectively, are high-lighted. In this case, the power counting is complicated, but can be set
up around the complex pole as~\cite{Gegelia:2016xcw}:
\begin{equation}
m_R-m_N \sim \varepsilon~,~~ m_R-m_\Delta \sim \varepsilon^2~,~~ m_\Delta-m_N \sim \varepsilon^2~,~~
M_\pi \sim \varepsilon^2~,
\end{equation}
where $\varepsilon$ denotes a small parameter. Again, let us calculate the self-energy to two loops
at the complex pole $z_R = m_R - i\Gamma_R/2$. By applying the cutting rules to these self-energy diagrams,
one obtains the graphs contributing to the decay amplitudes of the Roper resonance into the $\pi N$
and $\pi\pi N$ systems, leading to the total width
\begin{equation}
\Gamma_R = \Gamma_{R\to N\pi} +  \Gamma_{R\to N\pi\pi}~.
\end{equation}  
A somewhat lengthy calculation leads to:
\begin{eqnarray}
\Gamma(R\to N \pi) &=& 550(58) \, g_{\pi NR}^2\ {\rm MeV}~,\\ 
\Gamma(R\to N\pi\pi) &=& \Bigl(1.49(0.58)\,g_A^2 \,g_{\pi NR}^2-2.76(1.07)\, g_A^{} \, g_{\pi NR}^2\,g_R^{}\nonumber\\
&& + 1.48(0.58)\,g_{\pi NR}^2\, g_R^2 + 2.96(0.94)\,g_A^{}\, g_{\pi NR}^{} \,h_A^{} h_R^{} \nonumber\\
&& - 3.79(1.37)\,g_{\pi NR}^{}\,g_R^{} \,h_A^{} h_R^{} +9.93(5.45)\,h_A^2h_R^2\Bigr) \ {\rm MeV}~.
\end{eqnarray}
The total width thus depends on five LECs. The uncertainties in the round brackets are generated by the
uncertainties in the LECs. We use $g_A =1.27$ and $h_A=1.42\pm 0.02$. The latter value is the real
part of this coupling taken  from Ref.~\cite{Yao:2016vbz}.
As for the other unknown parameters,  the authors of~\cite{Gegelia:2016xcw} fixed $g_{\pi NR}$ so as to
reproduce the width  $\Gamma_{R\to \pi N}=(123.5\pm 19.0)$~MeV from the PDG. 
This yields $g_{\pi NR}=\pm (0.47\pm 0.04)$. In what follows, let us take the
positive sign for our central value and use the negative one as part of the
error budget. Further,  assume $g_R=g_A$ and $h_{R}=h_A$, the so-called maximal mixing assumption~\cite{Beane:2002ud}.
Then, one can make a prediction for the two-pion decay width of the Roper,
\begin{equation}
\Gamma (R\to N\pi\pi) = (41 \pm 22_{\rm LECs} \pm 17_{\rm h.o.})~{\rm MeV}~,
\end{equation}  
which is consistent with the PDG value of  ($67\pm 10$)~MeV. The error due to the neglect of the higher orders (h.o.)
is simply given by multiplying the $\varepsilon^5$ result (central value) with $\varepsilon = (m_R-m_N)/m_N
\simeq 0.43$. Clearly, to make further progress, we need an improved determination of the LECs $g_R$ and $h_R$.

\section{Lesson 3: New insights into  heavy-light meson spectroscopy}

In 2012, I had already  considered scattering the Goldstone boson octet $(\pi, K ,\eta)$  
off the $D$-meson triplet $(D^0, D^+, D_s^+)$. This  involves the
positive-parity  scalar charm-strange  mesons $D_{s0}^*(2317)$ and $D_{s1}(2460)$, which
are very narrow and are often interpreted as molecular $DK$ and $D^*K$ states, respectively.
Over the years, a number of puzzles has emerged that need to be answered: (1)
why are the masses of the $D_{s0}^\star(2317)$ and the $D_{s1}(2460)$ much lower than the
quark model predictions for $c\bar s$ mesons? (2) why is $M_{D_{s1}(2460)} - M_{D_{s0}^\star(2317)} \simeq M_{D^\star} -M_D$
within 2~MeV? and (3) why does one find $M_{D_{0}^\star(2400)} \gtrsim M_{D_{s0}^\star(2317)}$ and $M_{D_1(2430)} \simeq
M_{D_{s1}(2460)}$? Naively, one would expect that the mesons, where a light $(u,d)$ quark is substituted
by an $s$ quark, are heavier. Within the molecular picture, the combination of unitarized CHPT, lattice results
and precision data has led to a consistent picture, which I will only briefly discuss here. For more
details the reader should consult the comprehensive review~\cite{Guo:2017jvc}.

Concerning the first puzzle: It is resolved within the molecular scenario, as known since
long. This is further supported by recent lattice QCD calculations with $c\bar s$ and $DK$ interpolating
fields and relatively light pions~\cite{Mohler:2013rwa,Bali:2017pdv}. The $DK$ scattering length from
Ref.~\cite{Mohler:2013rwa}, $a_0 = -1.33(20)\,$fm, is consistent with the prediction of the molecular
scenario, $a_0^{\rm mol} = -2/\sqrt{\mu_{DK}\epsilon} \simeq -1\,$fm, with $\mu_{DK}$ the reduced mass
and $\epsilon \simeq 45\,$MeV the binding energy. Further, the quark mass dependence of the  $D_{s0}^\star(2317)$
calculated in Ref.~\cite{Bali:2017pdv} is perfectly consistent with the one in the molecular picture,
cf. Ref.~\cite{Du:2017ttu}. The wavefunction renormalization constant deduced from these results,
$1-Z = 1.04(0.08)(0.30)$ is also supporting the molecular scenario in which $Z^{\rm mol} = 0$. The second puzzle
is also naturally explained in this picture as a natural consequence of Heavy Quark Spin Symmetry (HQSS),
one simply substitutes the $D$-meson by a $D^*$-meson, see e.g.~\cite{Cleven:2010aw}. Using Heavy Quark
Flavor Symmetry (HQFS), one can furthermore easily make predictions for the
bottom analogues, $M_{B_{s0}^\star} \simeq M_B+M_K-45\,$MeV$ \simeq 5.730$~GeV and $M_{B_{s1}} \simeq M_{B^\star}+M_K
-45\,$MeV$ \simeq 5.776$~GeV, which are consistent with lattice QCD, $M_{B_{s0}^\star}^{\rm lat}  = (5.711 \pm 0.013
\pm 0.019)\,$GeV and $M_{B_{s1}}^{\rm lat}  = (5.750 \pm 0.017 \pm 0.019)\,$GeV~\cite{Lang:2015hza}. Finally,
let me resolve enigma (3).  Recently, the {\em Hadron Spectrum Collaboration} investigated coupled channel
$D\pi$, $D\eta$ and $D_{s}\bar{K}$ scattering with $I=1/2$ and found one pole corresponding to the
$D_0^\star(2400)$~\cite{Lang:2015hza}. These lattice energy levels were re-analyzed in Ref.~\cite{Albaladejo:2016lbb}
using unitarized CHPT and the values of the LECs already determined long before in~\cite{Liu:2012zya}.
They found two poles, completely analogous to the famous $\Lambda(1405)$, where the lighter one
has mass of about 2.1~GeV, thus resolving puzzle~(3). This pattern of two $I=1/2$ states emerges naturally
in the underlying formalism  since already the LO CHPT interactions are  attractive in two flavor multiplets,
to which the two states belong: the anti-triplet and the sextet~\cite{Kolomeitsev:2003ac,Albaladejo:2016lbb}. 
This double-pole structure had  already been noticed in earlier calculations,
see e.g.~\cite{Kolomeitsev:2003ac,Guo:2006fu} but only now finds
a natural explanation. Using HQSS and HQFS, one can extend this double-pole scenario to the $D_1, B_0^\star$ and $B_1$,
as summarized in Tab.~\ref{tab:double}.
\begin{table}[h]
\begin{center}
\begin{tabular}{|cccc|}
\hline
    & Lower Pol [MeV]  & Higher Pol [MeV]  & PDG [MeV]\\
\hline
 $D_0^\star$  & $\left(2105^{+6}_{-8}, 102^{+10}_{-11}\right)$  
             & $\left(2451^{+36}_{-26}, 134^{+7}_{-8}\right)$ 
             & {$(2318\pm 29, 134\pm 20)$}\\
 $D_1$       & $\left(2247^{+5}_{-6}, 107^{+11}_{-10}\right)$  
             & $\left(2555^{+47}_{-30}, 203^{+8}_{-9}\right)$ 
             & {$(2427\pm 40, 192^{+65}_{-55})$}\\
 $B_0^\star$  & $\left(5535^{+9}_{-11}, 113^{+15}_{-17}\right)$  
             & $\left(5852^{+16}_{-19}, 36\pm 5\right)$ 
             & {---}\\
 $B_1$       & $\left(5584^{+9}_{-11}, 119^{+14}_{-17}\right)$  
             & $\left(5912^{+15}_{-18}, 42^{+5}_{-4}\right)$ 
             & { ---}\\
\hline 
\end{tabular}
\caption{Two-pole scenario in various $I=1/2$ states in the heavy meson sector. Given are the mass $M$ and the
  half-width $\Gamma/2$ as $(M, \Gamma/2)$.}
\label{tab:double}
\end{center}
\vspace{-3mm}
\end{table}

\begin{figure}[t]
\includegraphics[width=4.75cm]{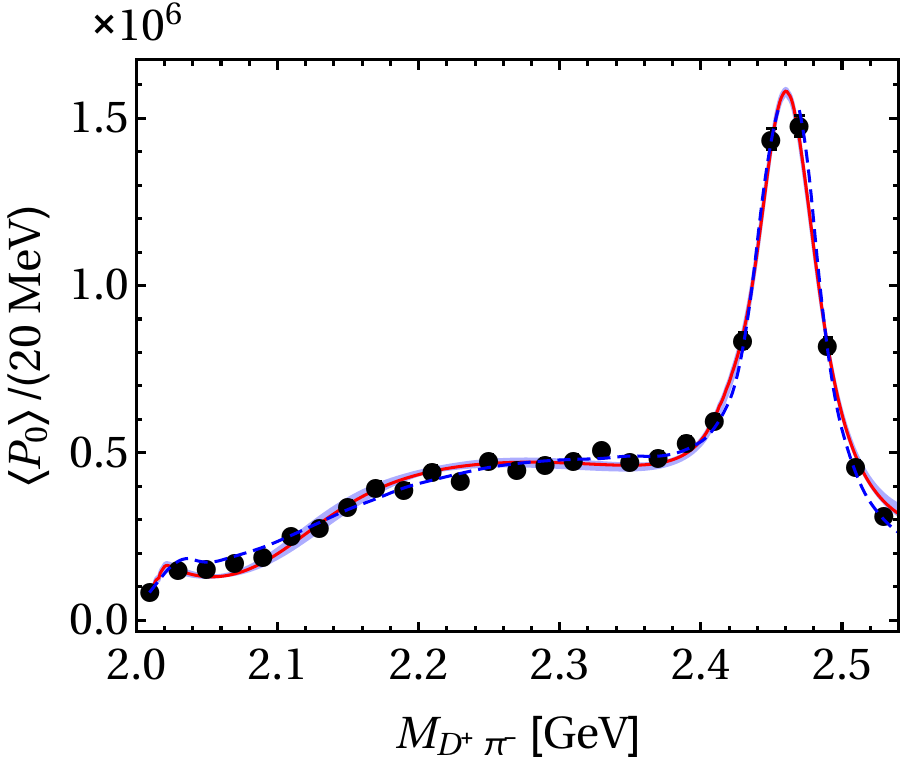}
\includegraphics[width=4.75cm]{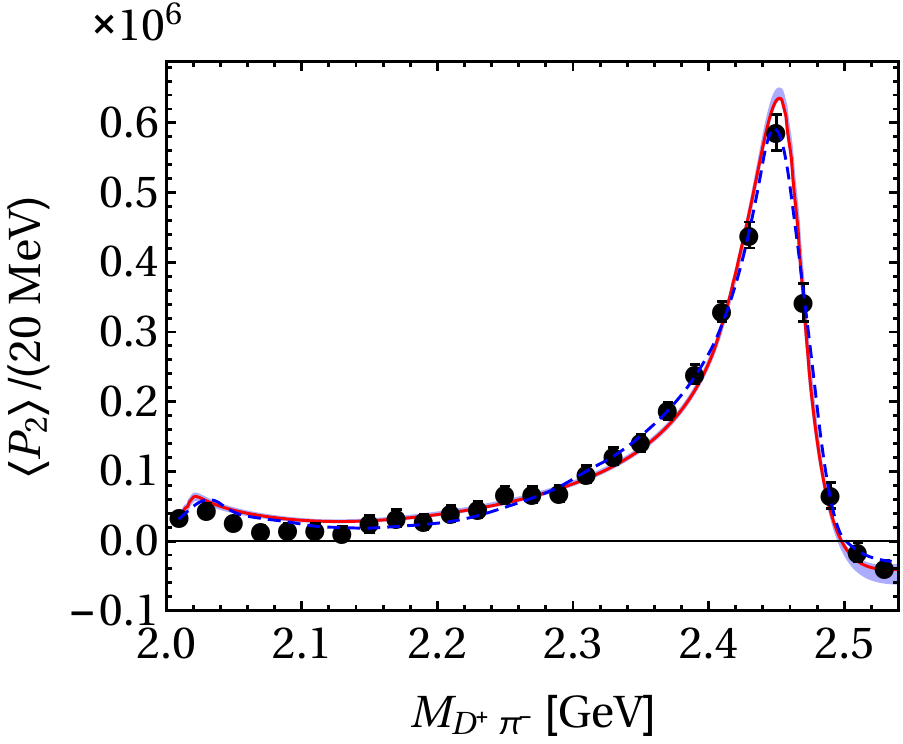}
\includegraphics[width=4.75cm]{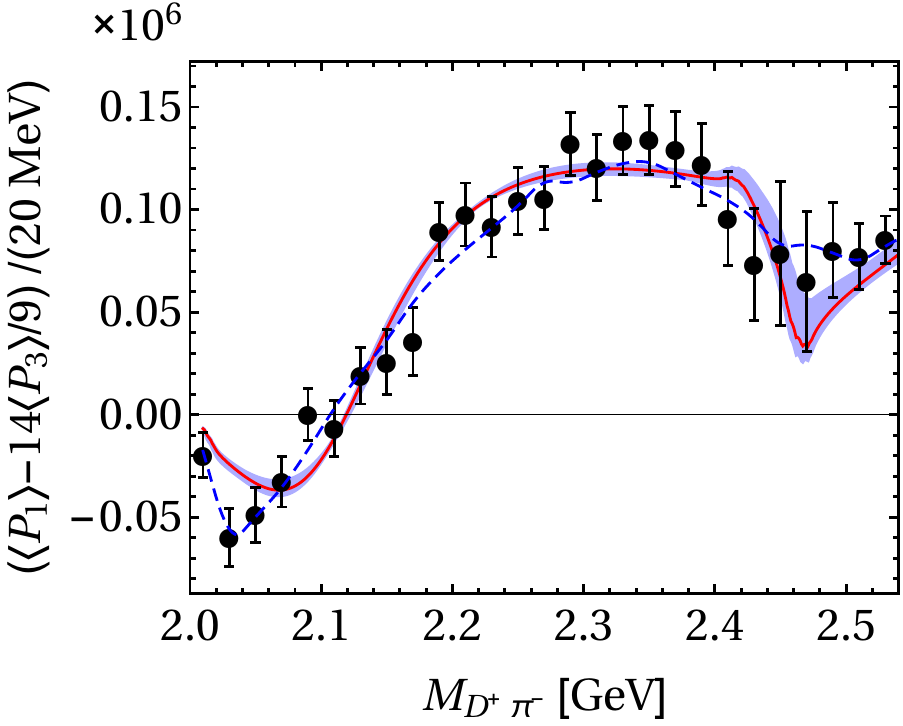}
\caption{Fit to the LHCb data for the angular moments $\langle P_0\rangle$,
  $\langle P_1\rangle - 14\langle P_3\rangle/9$ and $\langle P_2\rangle$ for the  $B^-\to
D^+\pi^-\pi^-$ reaction~\cite{Aaij:2016fma}. 
The largest error among $\langle P_1\rangle$ and $14\langle P_3\rangle/9$ in each bin is taken as
the error of $\langle P_1\rangle - 14\langle P_3\rangle/9$. The solid lines show  
the results of ~\cite{Du:2017zvv}, with error bands corresponding to the one-sigma uncertainties propagated 
from the input scattering amplitudes, while the
dashed lines stand for the LHCb fit using cubic splines for the $S$-wave.}
\label{fig:Dpi}
\vspace{-3mm}
\end{figure}

Remarkable, there is further support for this picture from the recent high precision results for
$B\to D\pi\pi$ from LHCb~\cite{Aaij:2016fma}. They determined accurately the so-called angular moments
from the $D\pi$ final-state interactions (FSI) that contain spectroscopic information. The corresponding
chiral Lagrangian for $\bar{B} \to D$ transition with the emission of two light pseudoscalars is known
since long~\cite{Savage:1989ub}. Combining it with the FSI from unitarized CHPT, one gets a two parameter
description of the corresponding amplitude in the energy region of the $D\pi$ system below 2.5~GeV,
${\mathcal A}(B^- \to D^+\pi^-\pi^-) = {\mathcal A}_0(s) +{\mathcal A}_1(s) + {\mathcal A}_2(s)$, in terms of
S-, P- and D-waves. In Ref.~\cite{Du:2017zvv} the same P- and D-waves as in the LHCb analysis were taken
and the S-wave was fixed form the coupled $D\pi, D\eta, D_s\bar K$ system already determined before. The
two parameters are one combination of LECs of the aforementioned chiral Lagrangian and the subtraction constant
in the two-meson loop functions $G_{ij}$, $\{ij\} \in \{D\pi, D\eta, D_s\bar K\}$. The angular moments are well
reproduced, see Fig.~\ref{fig:Dpi}. This picture can further be tested by confrontation with data, namely by
measuring the angular moments, in particular $\langle P_1\rangle -14\langle P_3\rangle/9$, with unprecedented
accuracy for the $B\to D^{(*)}\pi\pi$ and $B\to D_s^{(*)}\bar K\pi$ reactions. This can be done at LHCb and
Belle-II. One expects to see nontrivial cusp structures at the $D^{(*)}\eta$ and $D_s^{(*)}\bar K$ thresholds
in the former, and near-threshold enhancement in the $D_s^{(*)}\bar K$ spectrum in the latter~\cite{Albaladejo:2016lbb}.
Also, measuring the hadronic width of the $D_{s0}^*(2317)$, predicted to be about 100~keV in the
molecular scenario~\cite{Lutz:2007sk,Liu:2012zya}, while much smaller otherwise. This is a smoking-gun type of experiment
and will be performed by the $\overline{\text{P}}$ANDA experiment at FAIR. Finally, one can check  the
existence of the sextet
pole in lattice QCD with a relatively large SU(3) symmetric quark mass. And finally, one should search
for the predicted analogous bottom positive-parity mesons both experimentally and in lattice QCD.

I therefore conclude that the long accepted paradigm underlying open-flavor heavy
meson spectroscopy that identifies all ground states with $c\bar q$ or $b\bar q$
quark model states, is no longer tenable. In a broader view, the hadron spectrum must be viewed
as more than a  collection of quark model states, but rather as a manifestation of a more complex dynamics
that leads to an intricate pattern of various types of states that can only be understood by a
joint effort from experiment, lattice QCD and phenomenology.

\section{Lesson 4: A short remark on the pion-nucleon $\sigma$-term}

The situation concerning the Roy-Steiner analysis to precisely determine the pion-nucleon $\sigma$-term,
$\sigma_{\pi N}$, and the apparent tension with lattice determinations has been nicely reviewed by Martin
Hoferichter~\cite{MH}, see also the impressive work by the {\em BMWc Collaboration} reported by Laurent
Lellouch~\cite{LL}. One new piece of information concerns the flavor singlet term $\sigma_0$, related
to $\sigma_{\pi N}$ and the strangeness fraction $y$ via $\sigma_{\pi N} = \sigma_0 / (1-y)$. In
Ref.~\cite{Severt:2019sbz} the flavor decomposition of $\sigma_{\pi N}$ was re-analyzed in the framework of
baryon chiral perturbation to fourth order, employing both a covariant (EOMS) and the heavy baryon framework
and including also the low-lying decuplet (for the different approaches, see e.g. the review~\cite{Bernard:2007zu}).
Due to the tension between the lattice and Roy-Steiner determinations,
only continuum data were used, making the fits at fourth order difficult due to the large number of LECs.
Within some Bayesian approach, one arrives at the results collected in Tab.~\ref{tab:res1}.
\begin{table}[t]
\centering
\resizebox{15cm}{!}{
\begin{tabular}{|c|c|c|c|c|}
\hline
               &    HB           &   HB + decuplet  &   EOMS         &   EOMS + decuplet \\ 
\hline
$\Order(p^3)$  & 57.9(0.2)(17.0) & 88.6(0.2)(34.0)  & 46.4(0.2)(10.4)& 57.6(0.2)(17.0) \\ 
$\Order(p^4)$  & 64.1(31.7)(9.3) & 64.0(31.7)(18.7) & 51.8(31.4)(5.7)& 61.8(31.4)(9.3) \\
\hline
\end{tabular}
}
\caption{Results for $\sigma_0$ in MeV at third and fourth order. Here, ``+decuplet''
  means the inclusion of the decuplet, HB denotes the heavy baryon and EOMS the covariant
  approach. The first error comes from the uncertainties within the given order, the
  second error is an estimate of the neglected higher order effects following Ref.~\cite{Epelbaum:2014efa}.}
\label{tab:res1}
\end{table}
At third order, one finds a disturbingly large spread within the different approaches.
Similar results for the $\mathcal{O}(p^3)$ fits have been obtained earlier in \cite{Alarcon:2012nr}.  
The $\mathcal{O}(p^4)$ results for the covariant calculation only differ slightly from the
$\mathcal{O}(p^3)$ results. The fit without decuplet-resonances shifts more towards $\sigma_{\pi N}\simeq 59$~MeV,
while the fit with the decuplet is slightly above ${60}~{\rm MeV}$. The $\mathcal{O}(p^4)$ HB fit
results are both around ${64}~{\rm MeV}$. So the $\mathcal{O}(p^4)$ result for the HB approach
including decuplet baryons is much closer to $\sigma_{\pi N}$ than its corresponding $\mathcal{O}(p^3)$ result.
Given the central values for $\sigma_0$ from Tab.~\ref{tab:res1} and the precise value from
the Roy-Steiner analysis for $\sigma_{\pi N}$~\cite{Hoferichter:2015dsa}, we see that the strangeness
content $y \simeq 0$, but due to the large uncertainties, this can not be made more precise.

\section{Short summary \& outlook}

Picking up on my 2012 talk, it is obvious that hadron-hadron scattering continues to be a prime
playground for chiral dynamics. Over the years, a nice interplay of (U)CHPT, dispersion relations,
lattice QCD and experiment has developed, as discussed before in a few examples. There is a clear
challenge to lattice QCD, namely the precise determinations of the scattering lengths $a_0$ in $\pi\pi \to \pi \pi$,
$a_0^{1/2}, a_0^{3/2}$ in $\pi K\to \pi K$ and $a^+, a^-$ in $\pi N\to \pi N$. This would be premier tests
of the chiral QCD dynamics. Also, the latter are particularly important for another test of the lattice
QCD determinations of  $\sigma_{\pi N}$~\cite{Hoferichter:2016ocj}.

Furthermore, there has been
recent focus on heavy-light systems, where the marriage of CHPT with Heavy Quark Effective Theory and
dispersive techniques starts to pay large dividends. Also, (heavy) hadron decays play an ever increasing role,
largely due the large number of scattering processes with light mesons in the final states that are otherwise difficult
to access experimentally. On the continuum side, complex mass renormalization schemes and dispersion relations
appear most fruitful, while on the lattice side there is some focus on the development of
finite-volume formulas for three-body decays, see e.g. Ref.~\cite{Doring:2018xxx} (and references therein).
Clearly, we need $\omega \to 3\pi$ and $R\to N\pi\pi$ from
lattice QCD.

Finally, as nicely demonstrated by Evgeny Epelbaum in his talk~\cite{EE}, chiral symmetry
plays an indispensable role in nuclei. I look forward to many existing new results in the years to come.

\section*{Acknowledgements}

I thank all my collaborators for sharing their insight into the
topics discussed here.

%%%%%%%%%%%%%%%%%%%%%%%%%%%%%%%%%%%%%%%%%%%%%%%%%%%%%%%%%%%%%%%%%%%%%%%%%%%%%%%%%%%%%%%%%%%%%

\end{document}